\documentstyle[11pt,epsf]{article}

\def\fnote#1#2{\begingroup\def\thefootnote{#1}\footnote{#2}\addtocounter
{footnote}{-1}\endgroup}

\renewcommand{\thefootnote}{\fnsymbol{footnote}}

\def\beq{\begin{eqnarray}}
\def\eeq{\end{eqnarray}}
\def\bea{\begin{eqnarray*}}
\def\eea{\end{eqnarray*}}

\def\NPB#1#2#3{Nucl. Phys. {\bf B#1}, #3 (19#2)}
\def\PLB#1#2#3{Phys. Lett. {\bf B#1}, #3 (19#2)}

\def\PRP#1#2#3{Phys. Rep. {\bf #1}, #3 (19#2)}
\def\PRD#1#2#3{Phys. Rev. {\bf D#1}, #3 (19#2)}

\def\PRL#1#2#3{Phys. Rev. Lett. {\bf #1}, #3 (#2)}

\def\centeron#1#2{{\setbox0=\hbox{#1}\setbox1=\hbox{#2}\ifdim
\wd1>\wd0\kern.5\wd1\kern-.5\wd0\fi
\copy0\kern-.5\wd0\kern-.5\wd1\copy1\ifdim\wd0>\wd1
\kern.5\wd0\kern-.5\wd1\fi}}
\def\ltap{\;\centeron{\raise.35ex\hbox{$<$}}{\lower.65ex\hbox{$\sim$}}\;}
\def\gtap{\;\centeron{\raise.35ex\hbox{$>$}}{\lower.65ex\hbox{$\sim$}}\;}

\def\lsim{\mathrel{\ltap}}

\def\slashchar#1{\setbox0=\hbox{$#1$}           
   \dimen0=\wd0                                 
   \setbox1=\hbox{/} \dimen1=\wd1               
   \ifdim\dimen0>\dimen1                        
      \rlap{\hbox to \dimen0{\hfil/\hfil}}      
      #1                                        
   \else                                        
      \rlap{\hbox to \dimen1{\hfil$#1$\hfil}}   
      /                                         
   \fi}                                        %

\def\singleandthirdspaced{\baselineskip=\normalbaselineskip\multiply
    \baselineskip by 130\divide\baselineskip by 100}
\def\singlespaced{\baselineskip=\normalbaselineskip}

\newcommand{\newc}{\newcommand}
\newc{\Tr}{{\rm tr}}
\newc{\Det}{{\rm det}}
\newc{\Lambdaprime}{{\Lambda}}
\newc{\LambdaQCD}{\Lambda_{\rm eff}}
\newc{\mrat}{{\tilde m}}
\newc{\mratprime}{{\tilde m^\prime}}
\newc{\Log}{{\rm log}}
\newc{\Nc}{{N_c}}
\newc{\Nf}{{N_f}}
\newc{\aT}{{a_T}}
\newc{\aB}{{a_B}}
\newc{\bT}{{a_m}}
\newc{\dT}{{a_{M}}}
\newc{\cT}{c_T}
\newc{\mQ}{m_{\tilde Q}}
\newc{\mg}{m_{\tilde g}}
\newc{\msoft}{m_{\rm soft}}

\begin{document}

\begin{titlepage}
\begin{flushright}
{\large
hep-th/9801157 \\
SLAC-PUB-7739 
}
\end{flushright}

\vskip 1.2cm

\begin{center}

{\LARGE\bf Chiral symmetry breaking and effective lagrangians}

{\LARGE\bf for softly broken supersymmetric QCD}

\vskip 2cm

{\large
 Stephen P.~Martin$^a$
and James D.~Wells$^b$\fnote{\dagger}{Work supported by the Department of
Energy under contract DE-AC03-76SF00515. } 
} \\
\vskip 4pt
{\it $^a$Physics Department,
     University of California,
     Santa Cruz CA 95064  } \\
{\it $^b$Stanford Linear Accelerator Center,
     Stanford CA 94309} \\

\vskip 1.5cm

\begin{abstract}

We study supersymmetric QCD with $N_f<N_c$
in the limit of small supersymmetry-breaking masses and smaller quark
masses using the weak-coupling K\"ahler potential.  We calculate
the full spectrum of this theory, which manifests a chiral symmetry
breaking pattern similar to that caused by
the strong interactions of the standard model.
We derive the chiral effective lagrangian for the pion degrees of freedom,
and discuss the behavior in the formal limit of large squark and gluino
masses and for large $N_c$. We show that the resulting scalings of the
pion decay constant and pion masses in these limits differ from
those expected in ordinary nonsupersymmetric QCD. 
Although there is no weak
coupling expansion with $N_f=N_c$, we extend our results to this case
by constructing a superfield quantum modified constraint in the
presence of supersymmetry breaking. 

\end{abstract}

\end{center}

\vskip 1.0 cm

\end{titlepage}
\setcounter{footnote}{0}
\setcounter{page}{2}
\setcounter{section}{0}
\setcounter{subsection}{0}
\setcounter{subsubsection}{0}

\singleandthirdspaced

\section{Introduction}\label{sec:intro}
\setcounter{equation}{0}  
\indent

Solving strongly coupled field theories is hard.  Quantum chromodynamics
(QCD) provides a good example of the difficulties.  In the weak coupling
regime ($Q^2\gg \Lambda^2_{\rm QCD}$) we have gained much confidence over
the
last twenty years that QCD describes the interactions
between real quarks and gluons.  However, in the strong coupling regime
($Q^2\lsim \Lambda^2_{\rm QCD}$) the correspondence between QCD and nature
becomes more murky.  That is, QCD cannot be solved entirely in this energy
domain, in the sense that one cannot calculate analytically
from first principles the particle spectra,
the masses, and the dynamics of the low energy theory in
strong coupling.  Experimentally,
the independent left-right global chiral symmetries of
(nearly) massless QCD appear to be spontaneously broken to the vector
subgroup
by quark condensates.  Chiral perturbation theory incorporates these
symmetries and produces an effective lagrangian which accurately
describes interactions of the low-lying Nambu-Goldstone bosons (pions) 
of chiral symmetry breaking.  However, the effective 
chiral lagrangian is a phenomenological model not easily understood
from first principles.  Although much work continues to elucidate our
understanding of ordinary QCD in the strong coupling region, no 
entirely satisfactory mapping of the fundamental QCD lagrangian onto low
energy hadron physics has yet to be demonstrated.

In contrast to non-supersymmetric QCD, supersymmetric QCD (SQCD) 
holds out promise of understanding strong coupling 
more thoroughly.
A long struggle has culminated
in an impressive understanding of this supersymmetric theory.
The successes include knowledge of the exact 
$\beta$ function for the gauge coupling~\cite{NSVZ}, the
non-perturbative superpotential~\cite{ADS,TVY,amati,ils}
and a self-consistent description of the vacuum 
structure~\cite{seiberg:em,Peskin}.
The hope continues to be that we will gain important new insights
of field theory in general and QCD in particular
from the more controllable SQCD.  One step toward this
goal is to introduce supersymmetry breaking operators into
SQCD~\cite{MV,evans} and analyze the resulting theory.  This has been done
previously using canonical K\"ahler potentials
and small supersymmetry-breaking masses~\cite{ASPY,dhoker}. 
The spectrum of massless
states was calculated and shown to be in qualitative agreement
with QCD.  Work continues in an effort to find 
correspondence with 
QCD~\cite{alvarez-gaume,selipsky,farrar,sannino,Cheng}.

In this paper we will build upon previous analyses and calculate
the full mass spectra of supersymmetric QCD with $N_f\leq N_c$ using
the weakly coupled K\"ahler potential (for $N_f<N_c$), 
squark and gluino soft
supersymmetry-breaking masses ($m_{\tilde Q}$ and $m_{\tilde g}$, or
$m_{\rm soft}$ to represent either),  
and non-zero supersymmetry-preserving quark masses ($M_q$). 
We will then construct
the effective chiral lagrangian of the theory, and study the 
behavior of the pion masses and pion decay constant for $m_{\rm soft}
\gg M_q$. Furthermore, we will consider the problem of the decoupling
of the squarks and gluino by taking the formal limit $m_{\rm soft}
\gg \Lambda$ in these results.

\section{Softly-broken supersymmetric QCD with $\Nf<\Nc$}\label{sec:two}
\setcounter{equation}{0}  
\indent

Supersymmetric QCD is defined to be a supersymmetric theory
with an $SU(N_c)$ gauge group with $N_f$ quarks in the
fundamental ($Q$) and anti-fundamental ($\overline Q$) representations.
The chiral symmetry of SQCD with no explicit masses
is $SU(N_f)_L\times SU(N_f)_{R} \times U(1)_B \times U(1)_{R^\prime}$,
where $U(1)_B$ is baryon number and $U(1)_{R^\prime}$ is a non-anomalous
combination of the axial symmetry for the superfields and an $R$-symmetry
under which the gaugino fields transform non-trivially.
In this section we will be concerned with the $N_f<N_c$ case.
We denote the gauge-invariant meson field as $T_i^j=Q_i \overline Q^j$.
The non-perturbative superpotential for such a theory was
worked out by Affleck, Dine and Seiberg (ADS)~\cite{ADS} to be
\beq
W_{\rm ADS} = (\Nc - \Nf)\left( {\Lambda^{3\Nc - \Nf}\over
\Det T }\right )^{1\over \Nc-\Nf}.
\eeq
This superpotential preserves the chiral
symmetries mentioned above, and also respects the anomalous $R$-symmetry
if $\Lambda$ is taken to transform in the appropriate way \cite{Peskin}.

The vacuum of this theory corresponds to $\langle T\rangle$ running
to infinity.  Since $SU(N_c)$ with $N_f<N_c$ is asymptotically
free, we know that the weak-coupling K\"ahler potential is exact
for this theory for large $T$ along the $SU(N_c)$ flat directions.
The K\"ahler potential is then
\beq
K_T = 2\Tr [\sqrt{T^\dagger T}] .
\label{urKahler}
\eeq
There are two operational definitions of this K\"ahler potential.
First, one can write it as
\beq
K_T = 2 \sum_{i=1}^{N_f} \sqrt{\lambda_i},
\eeq
where $\lambda_i$ are the eigenvalues of the matrix $T^\dagger T$,
written in terms of the $N_f$ independent quantities invariant under
$SU(\Nf)_L \times
SU(\Nf)_R$ transformations: $\Det T^\dagger T$ and $\Tr[(T^\dagger T)^j]$
for
$j=1\ldots N_f-1$. For example, for $N_f=1$, the
meson
$T$ is just a single chiral superfield and the K\"ahler potential is
simply 
$K_T= 2 \sqrt{T^* T}$. For $N_f=2$, the K\"ahler potential can be written
explicitly as $K_T=2(\sqrt{\lambda_+} + \sqrt{\lambda_-})$, where
\beq
\lambda_{\pm} = {1\over 2}\left (
\Tr[T^\dagger T] \pm \sqrt{(\Tr[T^\dagger T])^2 - 4 \Det T^\dagger T}
\right ) .
\eeq
However, for general $N_f$ it is simpler in practice to write the K\"ahler
potential 
eq.~(\ref{urKahler}) as a power series expansion around the 
suspected minimum of the
scalar potential. This can be done in terms of
a new chiral superfield $Z$, by writing $T_i^j = t_0 (\delta_i^j +
Z_i^j)$, so that 
\beq
K 
= 
{t_0 \over 2}
 \Bigl (
\Tr [ Z^\dagger Z] -{1\over 4} \Tr [ Z^\dagger Z^2 + {Z^\dagger}^{2} Z
]
+
{1\over 16} \Tr [ 2 {Z^\dagger}^2 Z^2 - Z^\dagger Z Z^\dagger Z +
2 Z^\dagger Z^3 + 2{Z^\dagger}^3 Z ] + \cdots
\Bigr ) 
\label{Kexpand}
\eeq
(neglecting holomorphic pieces, which do not contribute to the
lagrangian), and
\beq
\Det T = t_0^{N_f} \Bigl ( 1 + \Tr[Z] + {1\over 2} \Tr[ Z]^2 -
{1\over 2} \Tr[Z^2] + \cdots \Bigr ) 
\eeq
for use in the superpotential.

We also want to incorporate soft masses into the lagrangian.  Introducing
soft squark masses $m_{\tilde Q}^2 (|\tilde Q|^2 + |\tilde{\overline Q}|^2)$ 
is
straightforward and corresponds to adding
$2m^2_{\tilde Q} \Tr [\sqrt{t^\dagger t}]$ to the scalar potential,
where $t_i^j$ is the scalar component of $T_i^j$. 
(For simplicity we take all
of the 2$N_f$ squark masses to be equal.) 
This quantity can once
again be defined either in terms of the square roots of the eigenvalues
of the matrix $t^\dagger t$ or by
its power series expansion:
\beq
\Tr [\sqrt{t^\dagger t}] = t_0 \left (N_c + {1\over 2} \Tr[ z+z^\dagger ]
- {1\over 8} \Tr[ (z-z^\dagger )^2] + \cdots \right ) ,
\label{mexpand}
\eeq
where $z_i^j$ is the scalar component of the superfield $Z_i^j$.
In order to include a gluino mass it is useful to include the
chiral superfield associated with the gluino bilinear:
$S= \frac{g^2}{32\pi^2}\lambda^\alpha \lambda_\alpha +\cdots$.
The superpotential with the $S$ field is~\cite{TVY}
\beq
W_S=S\left[ \Log\left( \frac{\Lambda^{3N_c-N_f}}{S^{N_c-N_f}\Det\, T}
 \right) +N_c-N_f\right] .
\label{Wtvy}
\eeq
Integrating out $S$ yields $W_{\rm ADS}$ and is equivalent to replacing 
$S=dW_{\rm ADS}/d\Log \Lambda^{3N_c-N_f}$~\cite{ils}.  Adding a gluino
soft mass is
accomplished simply by adding a term proportional to $\mg s$, where $s$
is the scalar component of $S$, before integrating out $S$.

In our discussion, we will also be considering a small amount
of explicit chiral symmetry breaking in the form of quark masses.
This can be accomplished with an additional term in the superpotential,
$W=W_{\rm ADS}+\Tr [M_q T]$. Throughout the following discussion we will
assume that the eigenvalues of the quark mass matrix $M_q$ are all
much smaller than the other scales in the problem, $\Lambda$ and
$m_{\rm soft}$, so that $M_q$ can be treated as a
perturbation.  

Incorporating all the elements described above, the full lagrangian
under consideration in the weak-coupling domain is
\beq
{\cal L} = 2\int d^4\theta\, \Tr [\sqrt{T^\dagger T}]
+\left(\int d^2\theta\, W + {\rm c.c.}\right) - V_{\rm soft};
\label{fulllagrangian}
\eeq
\beq
V_{\rm soft} = 
- c_T \mg \left( \frac{\Lambda^{3N_c-N_f}}{\Det\, t}
         \right)^{1\over {N_c-N_f}} 
+ {\rm c.c. } + 2 m^2_{\tilde Q} 
\Tr [\sqrt{t^\dagger t}],
\label{Vsoft}
\eeq
where $c_T=32\pi^2/g^2$.
Note that in the large $N_c$ limit~\cite{large Nc} 
one should think of $g^2N_c={\rm constant}$ and therefore $c_T\propto
N_c$.
Now it is straightforward to solve for the minimum of the potential
and quadratic fluctuations around it using the expansions in 
eqs.~(\ref{Kexpand})-(\ref{mexpand}).

In the limit of small $M_q$,
the minimum of the scalar potential occurs at 
$\langle t_i^j\rangle=t_0\delta_i^j$, where
\beq
\left ( {\Lambda^2\over t_0} \right )^{\Nc\over \Nc-\Nf}
= {1\over 2(N_c + N_f)\Lambda} \left [
c_T m_{\tilde g} + \sqrt{c^2_T m_{\tilde g}^2 + 4 (N_c^2-N_f^2) m_{\tilde Q}^2}
\right ] .
\eeq
For convenience we make the following definition
\beq
\mrat\equiv \Lambda \left ( {\Lambda^2\over t_0} \right )^{\Nc\over \Nc-\Nf}
=\left\{ \begin{array}{ll}
         m_{\tilde g}\frac{c_T}{N_c+N_f}, 
                & {\rm if}~m_{\tilde g}\gg m_{\tilde Q}; \\
         m_{\tilde Q}\sqrt{\frac{N_c-N_f}{N_c+N_f}}, 
                & {\rm if}~m_{\tilde Q}\gg m_{\tilde g},
	\end{array} \right.
\eeq
so that $\mrat \sim m_{\rm soft}$.
The $F$-term component of $T_i^j$ has dimensions of (mass)$^3$ and
gets a VEV equal to $F_T \delta_i^j$, where
\beq
F_T  = 2 \mrat^{N_f/N_c} \Lambda^{3-N_f/N_c} .
\label{defFT}
\eeq

The meson matrix can be parameterized as
\beq
t = t_0 {\rm exp}[(x+iy)/\sqrt{2\Nf} + \lambda^a (x_a + i y_a)]
\eeq
where $\langle t_i^j \rangle = t_0 \delta_i^j$ is the meson VEV
given above, and $\lambda^a$ 
(with $a=1,\ldots,$$N_f^2-1$) are the generators of $SU(N_f)$, normalized 
to $\Tr[\lambda^a \lambda^b] = {1\over 2} \delta^{ab}$.
The fields $x$, $y$, $x_a$ and $y_a$ 
are dimensionless. The corresponding meson fields
with canonically-normalized kinetic terms, 
$V$, $A$, $V_{a}$ and $\pi_{a}$, are obtained
by multiplying each of these fields by $F_\pi$, given by
\beq
F^2_{\pi} = {t_0/2}. 
\label{defFpi}
\eeq

The full spectrum of the model contains the flavor-adjoint pions ($\pi_a$),
the heavy scalars
($V$, $V_a$), heavy pseudo-scalar ($A$),
and heavy fermions ($\psi_0$ and $\psi_a$). For the
masses of the heavy particles, we find
\beq
m^2_{V} &=& {4\Nc\over \Nc-\Nf} m_{\tilde Q}^2+ 
           {2\Nc\over (\Nc-\Nf)^2} c_T m_{\tilde g}\mrat ,
\label{heavyV}
\\
m^2_{V_{a}} &=& {4\Nc\over \Nc+\Nf} m_{\tilde Q}^2
          + {2 \over \Nc+\Nf} c_T m_{\tilde g}\mrat ,\\
m^2_{A} &=&  {2\Nf\over (\Nc-\Nf)^2}c_T m_{\tilde g} \mrat , 
\label{heavyA}
\\
m_{\psi_0} &=& {\Nc+\Nf\over \Nc-\Nf}\mrat ,
\label{heavypsi}
\\
m_{\psi_a} &=& \mrat , 
\label{heavypsia}
\eeq
 again neglecting $M_q\ll\Lambda, m_{\rm soft}$.  
The pions $\pi_a$ are massless when $M_q=0$, and (as in ordinary QCD) 
obtain a (mass)$^2$ matrix which is linear in the quark mass
matrix $M_q$. We find
\beq
(m^2_{\pi})_{ab} = 8 \Tr[M_q \lambda^a \lambda^b] \,\mrat.
\label{pionmass}
\eeq  

In the limit $M_q\rightarrow 0$ there is
an interesting correspondence with ordinary QCD
as originally pointed out in~\cite{ASPY}. Light pions
are indicative of the Nambu-Goldstone modes of the $SU(N_f)_L\times 
SU(N_f)_R$ chiral symmetry spontaneously breaking to its vector subgroup.
Also, when the gluino mass is zero there is an extra Nambu-Goldstone mode
associated with the spontaneous breaking of the non-anomalous
$U(1)_{R^\prime}$
symmetry. This massless state is $A$, and it only gains mass when
$U(1)_{R^\prime}$
is explicitly broken by a soft gluino mass. 
All of the masses scale like $N_c^0$ at large $N_c$, except $m^2_A$ which 
scales like $1/N_c$ (recall that $c_T\propto N_c$).  
This is because in the 
large $N_c$ limit~\cite{large Nc} 
the chiral symmetries of SQCD are promoted to
$U(N_f)_L\times U(N_f)_R$, leaving an extra Nambu-Goldstone boson
in the spectrum when it is spontaneously broken to $U(N_f)_V$.  
Thus, the $A$ field
in supersymmetric QCD can be identified with the $\eta'$ analog 
in ordinary QCD.

\section{The chiral effective
lagrangian and scaling behavior towards the decoupling
limit}\label{sec:three}
\setcounter{equation}{0}  

\indent

In the weakly-coupled limit of supersymmetric QCD discussed in the
previous section, we find that the chiral effective lagrangian for the
pion degrees
of freedom in the limit $M_q \ll m_{\rm soft}$ can be
written as 
\beq
\label{chiral}
{\cal L} = 
F_\pi^2 \Tr[\partial^\mu U^\dagger \partial_\mu U]
+ F_T (\Tr[U M_q] + {\rm c.c.})
\label{chirallagrangian}
\eeq
where $F_\pi$ and $F_T$ are given in eqs.~(\ref{defFpi}) and
(\ref{defFT}), and $U =
{\rm exp}[i\lambda^a\pi_a/F_\pi]$. 
This form is familiar as the usual chiral effective lagrangian used
to describe the effective low-energy dynamics of pseudo-scalar mesons in
ordinary QCD. Thus it is clear that $F_\pi$ should be interpreted as the
pion decay constant.
The pion (mass)$^2$ matrix is given simply in terms of $F_T$ and $F_\pi$
by 
\beq
(m^2_\pi)_{ab}=2\Tr[M_q \lambda^a \lambda^b] \frac{F_T}{F_\pi^2}
\eeq
[cf.~eqs.~(\ref{defFT}),
(\ref{defFpi}) and
(\ref{pionmass})].
It is tempting to associate $F_T$ with the quark bilinear condensate
$\langle \psi_Q \psi_{\overline Q} \rangle$~\cite{ASPY} in accordance
with our picture of chiral symmetry breaking in ordinary QCD.  
Because the confining phase and the Higgs phase of the theory are
smoothly connected to each other \cite{conHiggs}, this correspondence
suggests that the chiral effective lagrangian for the pion degrees of
freedom might be obtained as a sensible limit of the weak-coupling
SQCD lagrangian.
However, since
$F_T= -\psi_Q\psi_{\overline Q}+QF_{\overline Q}+\overline QF_{Q}$ in terms
of
the original quark fields, it is not entirely
clear that $\langle F_T\rangle$ can be identified solely as a
condensate of the original quark fermions in the weak-coupling regime,
and in the strong-coupling regime the calculation with
the K\"ahler potential eq.~(\ref{urKahler}) is unreliable.

To investigate this correspondence further,
consider the scaling of the chiral effective
lagrangian parameters $F_\pi$ and $F_T$ with $m_{\rm soft}$.
In the regime
$\mrat \sim m_{\rm soft} \ll\Lambda$ where weak coupling makes our
calculations
reliable, we find from the results of the previous section that  
\beq
F_\pi^2& \propto & \left ({\mrat\over \Lambda}\right )^{-{(N_c-N_f)
\over N_c}}
\Lambda^{2} ,
\label{weakscalingFpi}\\
F_T & \propto & \left ({\mrat\over \Lambda }\right )^{{N_f \over N_c}}
\Lambda^{3}
.
\label{weakscalingFT} 
\eeq
Note that as $m_{\rm soft}$ increases, $F_\pi^2$ decreases but
$F_T$ and $m^2_\pi$ grow.

To compare the scaling behavior of these quantities
in the formal limit of large $\msoft$ with the behavior expected in
ordinary QCD, we must determine how the SQCD scale $\Lambda$ is related to
the effective QCD scale $\LambdaQCD$ after the squarks and gauginos are
decoupled. Using one-loop renormalization group matching, one finds that
\beq
\LambdaQCD = \Lambda \left (m_{\tilde g}\over \Lambda \right )^{2N_c\over
11 N_c - 2 N_f} \left (m_{\tilde Q}\over \Lambda \right )^{N_f\over
11 N_c - 2 N_f} \label{decoupleboth}
\eeq
when $m_{\tilde g}, m_{\tilde Q} \gg \Lambda$, or
\beq
\LambdaQCD = \Lambda
\left ( {m_{\tilde g} \over \Lambda } \right )^{2\Nc \over 11 \Nc-\Nf}
\label{decouplegluino}
\eeq
if only the gluino is heavy compared to $\Lambda$, or
\beq
\LambdaQCD = \Lambda
\left ( {m_{\tilde Q} \over \Lambda } \right )^{\Nf \over 9 \Nc-2\Nf}
\label{decouplesquarks}
\eeq
if only the squarks are heavy compared to $\Lambda$.
Since the chiral effective lagrangian parameters for ordinary QCD (in the
limit
that the squarks and gluinos decouple) should only depend on the
soft masses through the scale $\Lambda_{\rm eff}$, one expects that 
the pion dynamics would be governed by a lagrangian with the same form
as eq.~(\ref{chirallagrangian}), but with
\beq
F_\pi^2 &\propto& \LambdaQCD^2
\label{expectedscalingFpi}\\
F_T &\propto& \LambdaQCD^3
\label{expectedscalingFT}
\eeq
so that
$m_\pi^2 \propto \LambdaQCD M_q$. 
Unfortunately, it is easy to
check that eqs.~(\ref{expectedscalingFpi}) and
(\ref{expectedscalingFT})
do not have the same scaling behavior with $m_{\rm soft}$ in the formal
decoupling limit as the weak-coupling results 
eqs.~(\ref{weakscalingFpi}) and (\ref{weakscalingFT}), no matter
which of eqs.~(\ref{decoupleboth}), (\ref{decouplegluino}) or
(\ref{decouplesquarks}) applies.
Furthermore, the
large $N_c$ scaling of the large-$\msoft$ chiral lagrangian
does not conform with expectations from ordinary QCD. 
Non-supersymmetric chiral perturbation
theory implies~\cite{large Nc} 
that $F_\pi^2\propto N_c$ and $F_T\propto N_c$,
but the formal decoupling limit of the
weakly-coupled SQCD chiral lagrangian scales as
$F_\pi^2\propto N_c^0$ and $F_T\propto N_c^0$.

The failure of the weak-coupling results found in section 2 to
agree with 
ordinary QCD is not unexpected.
For one thing, one should probably allow for the possibility of a much
more general scaling of the
K\"ahler potential with $T$ as one approaches the large-$m_{\rm soft}$
regime. The lack of non-renormalization theorems
for the K\"ahler potential means that 
other contributions, over which we have little control, 
likely will dominate for the small $T$
region relevant for the decoupling to ordinary QCD.
One such contribution can be seen by considering the original
K\"ahler potential for the composite glueball chiral superfield $S$, of
the form 
$K_S \sim (S^* S)^{1/3}$ \cite{TVY}. Integrating out $S$ using this
K\"ahler
potential and the superpotential eq.~(\ref{Wtvy}), one finds an
additional
contribution to the K\"ahler potential for the remaining $T$ fields of the
form
\beq
K_S \sim (S^*S)^{1/3}\sim 
   \Lambda^2 \left(\frac{\Lambda^{4N_f}}{\Det [T^\dagger T]}
\right)^{1/3(N_c-N_f)}.
\label{SKahler}
\eeq
In weak coupling the neglect of this term in the K\"ahler potential is
justified because $K_T=2\Tr [\sqrt{T^\dagger T}]$
always dominates for large $T$.  However, as we move towards strong
coupling this term becomes important, as do presumably many other
contributions which are singular for small $T$. Moreover, there are
additional unknown contributions coming from integrating out the entire
spectrum of composite massive hadronic degrees of freedom.
Likewise, the soft terms can be more complicated as well in the strong
coupling
region~\cite{dhoker,sannino}, 
leading to different scaling behavior of the chiral lagrangian.

In ref.~\cite{ASPY}, a canonical K\"ahler potential and soft terms
were used, with arbitrary prefactors which parameterize our lack of
knowledge of the effects just mentioned. 
Alternatively, we can introduce such renormalization constants for the
K\"ahler
potential, soft terms and quark masses in the weak-coupling lagrangian:
\beq
K = 2a_T  \Tr[\sqrt{T^\dagger T}]
\label{canKahler}
\eeq
instead of eq.~(\ref{urKahler}); 
\beq
V_{\rm soft} = 
- c_T \mg \left( \frac{\Lambda^{3N_c-N_f}}{\Det\, t}
         \right)^{1\over {N_c-N_f}} 
+ {\rm c.c.} + 2{a_m} m^2_{\tilde Q} \Tr [\sqrt{t^\dagger t}]
\eeq
instead of eq.~(\ref{Vsoft}). In the presence of supersymmetry breaking,
one must also consider a similar renormalization of the quark mass
terms $M_q \rightarrow \dT
M_q$. Here $a_T$,
$\bT$, $c_T$ and $\dT$ are
dimensionless quantities which can {\it a priori} have an arbitrary
dependence on $\Lambda$ and the soft terms in the underlying lagrangian,
which in an effective lagrangian approach can also mimic any dependence
of these terms on the dynamical fields.
The limits $a_T, a_m, a_M \rightarrow 1$ can of course be used to recover
the weak coupling results.

Using this as a toy model, one can show that the VEV occurs for $\langle
t_i^j \rangle  = t_0 \delta_i^j$, with $t_0$ satisfying
\beq
\left ( {\Lambdaprime^2\over t_0} \right )^{\Nc\over \Nc-\Nf}
= {a_T\over 2(N_c + N_f) \Lambdaprime} \left [
c_T m_{\tilde g} + \sqrt{c^2_T m_{\tilde g}^2 + 4 (N_c^2-N_f^2)
\bT m_{\tilde
Q}^2/a_T}
\right ] \equiv \frac{\mratprime}{\Lambdaprime}
\label{canVEV}
\eeq
in the limit of small $M_q$. In terms of $\mratprime$,
the masses of the heavy mesons and fermions
are given  in this case by eqs.~(\ref{heavyV})-(\ref{heavypsia}) with the
replacements $m_{\tilde Q}^2 \rightarrow \bT m_{\tilde Q}^2/a_T$
and $\mrat \rightarrow \mratprime/a_T$.
The pion masses and interactions in the low energy effective theory
are given by a lagrangian of exactly the same form as
eq.~(\ref{chirallagrangian}), but now with $M_q \rightarrow \dT M_q$ and
\beq
F_\pi^2 &=&  
{a_T \over 2} \left ({\mratprime \over \Lambdaprime}\right
)^{-{(N_c-N_f)\over N_c}}
\Lambdaprime^2
 \label{Fpiaspy}\\
F_T &=& {2\over a_T}
\left ({\mratprime \over \Lambdaprime}\right )^{{N_f \over N_c}}
\Lambdaprime^{3} .
\label{FTaspy}
\eeq
If one attempts to make these results agree with the expected
behavior in the decoupling limit
eqs.~(\ref{decoupleboth})-(\ref{expectedscalingFT}), then one finds that
\beq
a_T & \sim & 
\left ( {\Lambda_{\rm eff} \over \Lambdaprime} \right )^2
\left ( {\mratprime \over \Lambdaprime} \right )^{{N_c - N_f \over N_c}}
\\
\dT & \sim & 
\left ( {\Lambda_{\rm eff} \over \Lambdaprime} \right )^5
\left ( {\mratprime \over \Lambdaprime} \right )^{{N_c - 2 N_f \over N_c}}
.
\eeq
However, it is hard to imagine a sound origin for these assignments, in
view of our lack of knowledge of the lagrangian for small $T$.
 Matching on to a correct quantitative
description of
the decoupling of superpartners requires the inclusion of non-trivial
K\"ahler potential terms and
terms containing higher superderivatives (giving rise to operators which
contain more than two powers of the auxiliary fields) over which we do not
seem to have even qualitative control.
Thus it appears that the difficulty of finding a 
softly broken supersymmetric QCD lagrangian which gives the
correct chiral lagrangian is perhaps equally as difficult as solving
ordinary QCD.  The language of supersymmetry does not seem to have
made this task easier.

\section{Supersymmetric QCD with $N_c=N_f$}\label{sec:four}
\setcounter{equation}{0}  

\indent

When the number of flavors is equal to or greater than 
the number of colors there are non-trivial  gauge-invariant
baryonic degrees of freedom.  In the $N_c=N_f$
case, two such baryon fields (implicitly antisymmetric in color) are
allowed:
\beq
B & = & Q_1Q_2\cdots Q_{N_c} , \\
\overline B & = & \overline Q_1\overline Q_2\cdots \overline
Q_{N_c}.
\eeq
Decoupling flavors to form an $N_f<N_c$ theory consistent
with the ADS superpotential requires that the meson fields and
the baryon fields satisfy a quantum modified constraint~\cite{seiberg:em}:
\beq
\Det T -B\overline B = \Lambda^{2N_c}.
\eeq
Any vacuum which resembles ordinary QCD requires $\langle B\rangle=
\langle \overline B\rangle =0$ to preserve baryon number and
$\langle T_i^j\rangle \propto \delta_i^j$ to preserve the vectorial
$SU(N_f)$. This implies
$\langle T_i^j\rangle=\delta_i^j\Lambda^2$ from the quantum modified
constraint.

An unfortunate corollary to the quantum modified constraint is the
inability to have even approximate control over the K\"ahler potential
for any values of the soft terms. In the $N_f<N_c$
case we were able to have confidence in the small supersymmetry breaking
results because the theory was still at weak coupling.
However, now the vacuum at $\langle
T_i^j\rangle=\delta_i^j\Lambda^{2}$
corresponds to strong coupling.  With no guiding principle
for the K\"ahler potential we choose the weak-coupling K\"ahler potential
for $T_i^j$ and a canonical K\"ahler potential for $B$ and $\overline B$:
\beq
K = 2\aT \Tr[\sqrt{T^\dagger T}] + {\aB\over \Lambda^{2\Nc-2}}
(B^* B + \overline B^* \overline B).
\eeq
(Other K\"ahler potentials, including a canonical K\"ahler potential for
$T$,  give qualitatively similar results.)
We also choose soft terms of the form
\beq
V_{\rm soft}= - (c_T m_{\tilde g} s + {\rm c.c.}) +
2 \bT m^2_{\tilde Q} {\Tr}[\sqrt{t^\dagger t}]+{m^2_{\tilde B}
\over \Lambda^{2\Nc-2}}
(|b|^2+|\overline b|^2).
\eeq
where $m_{\tilde B}^2$ originates from the soft
terms in the underlying lagrangian in some undetermined way,
and $b$ and $\overline b$ are the scalar components of $B$ and
$\overline B$. Based on the constituent squark description
of $T$, $B$ and $\bar B$, one might naively expect that 
$m_{\tilde B}^2\simeq N_c^2 m^2_{\tilde Q}$.  However, 
soft terms for the baryons in
the confining description have no direct analog to the soft terms
of squarks in the Higgs description.  Therefore, if the $N_c=N_f$ case had
a self-consistent weak coupling expansion then it would appear 
that the baryons should have zero soft masses.  
Since no weak coupling
domain exists for this theory, and given the above ambiguities it
is best to treat $m_{\tilde B}$ as a free parameter.

The $S$-dependent superpotential can be written as~\cite{TVY}
\beq
W_S = S \, {\Log}\left ({\Det T - B \overline B\over \Lambda^{2\Nc}}
\right
) + a_M \Tr[M_q T].
\eeq
We ignore the K\"ahler potential for the $S$ term since it would 
induce more complicated K\"ahler terms for $T$, $B$ and $\overline B$
after it is integrated out; these can hopefully be absorbed into the
definitions of $a_T$ and $a_B$ in an effective field theory approach.  
So, treating $S$ as a Lagrange multiplier, 
the superpotential can be written simply as
$W=a_M \Tr [M_q T]$ subject to the new superfield constraint
\beq
\Det T - B \overline B = 
\Lambda^{2\Nc} 
(1-\theta\theta c_T m_{\tilde g}).
\label{superconstraint}
\eeq
Integrating out the auxiliary fields for $T_i^j$, $B$ and 
$\overline B$
subject to this constraint produces a scalar
potential (neglecting $M_q$ for now):
\beq
V &=& -{ a_B a_T c_T^2 m^2_{\tilde g}\Lambda^{4N_c} \over
     2 a_B \prod_i |t_i|^2 \sum_j |t_j|^{-1} \,+\, a_T \Lambda^{2N_c-2}
(|b|^2 + |\overline b|^2) } \nonumber
\\ &&+2a_m m^2_{\tilde Q} \sum_i |t_i|
+{m^2_{\tilde B}\over \Lambda^{2\Nc-2}}
(|b|^2+|\overline b|^2)
\label{Vugly}
\eeq
where $t_i$ are the eigenvalues of $t_i^j$,
subject to the scalar field constraint 
$\prod_i t_i = b\overline b + \Lambda^{2\Nc}$. 
The minimum of this potential always occurs for $t_i^j = t_0 \delta_i^j$
and $b =-\overline b$.

A vacuum like that of ordinary QCD with conserved baryon number and
chiral symmetry breaking
corresponds to $t_0 = \Lambda^2$ and $b = \overline b = 0$.
This is a local minimum of the scalar potential if
\beq
{c_T^2 m_{\tilde g}^2 \over N_c^2} \left (a_T -(2N_c-1) a_B 
\right ) > 4 {a_B\over a_T} (\bT m_{\tilde Q}^2 - m_{\tilde B}^2).
\label{localmin}
\eeq
Note that in the limit $a_B \ll a_T$, this condition is always satisfied.
A necessary but not sufficient
condition for this to also be a global minimum of the potential is
\beq
{c_T^2 m_{\tilde g}^2 \over N_c^2} \left (a_T - N_c a_B 
\right ) > 4 (\bT m_{\tilde Q}^2 - {m_{\tilde B}^2 \over N_c}).
\label{globalmin}
\eeq
This is because there is always a local minimum
at $t_0=0$ with $b = -\overline b =  \Lambda^{N_c}$, which has lower
energy
than the $b = \overline b=0$ minimum if eq.~(\ref{globalmin}) is not
satisfied. Note also that for some values of the parameters, the global
minimum of the potential can occur for both $b=-\overline b $ and $t_0$
non-zero, if the $m_{\tilde g}^2$ term dominates and $a_T < a_B (2N_c-1)$. 
If so, then
both baryon number and chiral symmetry will be broken in the vacuum state.
However, one must remember that since the description of the theory in
terms of the parameters
$a_T$, $a_m$, $a_M$, $m_{\tilde B}^2$ and $c_T$ is only an
effective one, it is not clear that the global properties of the potential
are significant. Therefore, presumably only the condition
eq.~(\ref{localmin}) should be considered as a constraint on the
parameters of the model, even if baryon number is assumed or required to
be unbroken.

The spectrum of this model in the baryon number conserving vacuum contains
heavy flavor-adjoint scalars 
$V_a$ with 
\beq
m_{V_a}^2 = {c_T^2 m_{\tilde g}^2 \over 2N_c^2} + {2 \bT m_{\tilde Q}^2
\over a_T}
\eeq 
and heavy flavor-adjoint fermions with
\beq
m_{\psi_a} = {c_T m_{\tilde g} \over 2 N_c} .
\eeq
Note that these results agree with those for the corresponding states in
the $N_f < N_c$ case as found above, if one takes the $N_f \rightarrow N_c$
limit. The flavor-singlet components of
$T$ and $\psi_T$ are removed by the scalar and fermionic components of the
superfield constraint (\ref{superconstraint}). (The latter constraint
takes the form $\Det T \, \Tr[T^{-1} \psi_T] - B \psi_{\overline B} -
{\overline B}
\psi_{B} = 0$.) This corresponds to the singularity in 
eqs.~(\ref{heavyV}), (\ref{heavyA}) and (\ref{heavypsi}) as $N_f
\rightarrow N_c$.
Instead, there are some additional
degrees of freedom corresponding to the baryons
in the $N_c=N_f$ case which have no direct analog
in the $N_f<N_c$ theory.  The complex scalars $B$ and $\overline B$
mix to form two degenerate pairs of real scalar mass eigenstates,
with
\beq
m^2_{B\overline B} =
\left ( 
{a_T^2 c_T^2 m_{\tilde g}^2 \over 4 N_c^2 a_B^2}
+ {m_{\tilde B}^2 \over a_B}
\right ) 
\pm
\left ( 
{a_T c_T^2 m_{\tilde g}^2 (2N_c-1)\over 4 N_c^2 a_B}
+ {\bT m_{\tilde Q}^2 \over a_B}
\right ).
\eeq
The two heavy fermionic partners of $B$ and $\overline B$ pair up to get
a Dirac mass,
\beq
m_{\psi_B\psi_{\overline B}} = {a_T c_T m_{\tilde g} \over 4 N_c a_B}.
\eeq
(In all of the preceding, we have neglected $M_q \ll m_{\rm soft}$ as
before.)

Of particular interest is the chiral lagrangian governing the
pion masses.  It has the same form as eq.~(\ref{chiral}), with now
$M_q \rightarrow a_M M_q$ and
\beq
F_\pi^2 &=& {a_T \Lambda^2 \over 2},
\label{fpinc}\\
F_T &=& {c_T m_{\tilde g}\Lambda^2 \over \Nc},
\eeq
and
\beq
(m_{\pi}^2)_{ab} = 2 a_M \Tr[M_q \lambda^a \lambda^b] {F_T \over
F_\pi^2}
=
4 \Tr[M_q \lambda^a \lambda^b] {a_M c_T m_{\tilde g}
\over \aT \Nc} .
\label{mpinc}
\eeq
If $\mg$ vanishes, then the pions are massless in linear order in $M_q$.
This can be understood easily from the fact that with $m_{\tilde g} = 0$,
the scalar potential has only a quadratic dependence
on $M_q$. Equations (\ref{fpinc})-(\ref{mpinc}) again agree
exactly with the
$N_f \rightarrow N_c$ limit of the corresponding results found above
for $N_f < N_c$.  If one could make a weak-coupling 
expansion
with $a_T, a_m, a_M \rightarrow 1$, then
the $N_c=N_f$ case would experience the same
similarities (chiral symmetry breaking pattern) and the same
dissimilarities ($m_{\rm soft}$ and $N_c$ scaling in the chiral
lagrangian) as the $N_f<N_c$ case when comparisons are made to
ordinary nonsupersymmetric QCD. It is possible to choose values for 
the unknown parameters so that the expected scalings for the decoupling
limit of ordinary QCD are realized; however, as noted for $N_f < N_c$,
these
assignments are hard to justify with actual calculations. 

\section{Discussion}\label{sec:conclusion}
\setcounter{equation}{0}  

\indent

We have analyzed the spectroscopy and chiral lagrangian of
supersymmetric QCD with $N_f\leq N_c$ in an attempt to learn
more about these theories with small explicit chiral symmetry
breaking and small supersymmetry breaking.  For $N_f<N_c$ the
supersymmetric theory is weakly coupled in this limit, and the
chiral symmetry breaking and spectroscopy are reliably calculated.
Our calculations have been carried out using confining phase degrees
of freedom.  We can compare our results with those of ref.~\cite{dhoker}
which worked with Higgs-phase degrees of freedom in the limit of
$M_q=0$ and $m_{\tilde g}=0$. The particle spectrum we find in
this limit from the lagrangian
of eq.~(\ref{fulllagrangian}) agrees with their spectrum,
thus demonstrating equivalence when
the appropriate weak-coupling K\"ahler potential and soft terms
are used in both approaches.
We have presented results also for the chiral effective lagrangian 
parameters in this theory.

For $N_f \geq N_c$, the situation becomes less controlled, because the
theory has no
reliable weak coupling limit with
the same gauge group for small supersymmetry breaking masses.
Nevertheless, we have
calculated the potential, chiral lagrangian and spectrum 
assuming a modified weak-coupling K\"ahler potential for $N_f = N_c$,
where one can at least obtain the ordinary QCD-like chiral symmetry
breaking pattern. 
We also noted the possible existence of an exotic phase with both
broken chiral symmetry and broken baryon number, which may exist for
finite squark and gluino masses, in addition to the possible vacuum with
spontaneously broken baryon number and conserved chiral symmetry
as argued in \cite{ASPY}.
Unfortunately, the effective lagrangian approach
is quite incapable of answering whether the true softly broken SQCD 
theory can actually have
these vacuum states. If they do exist at all, they cannot occur for
arbitrarily large squark masses, because QCD with only
vectorlike fermions cannot spontaneously break baryon number \cite{VW}.

We found that attempts to match the chiral effective lagrangian for
softly-broken SQCD to ordinary QCD 
by merely extrapolating the weak coupling results to the formal limit of
large squark and gluino masses
do not yield the correct scaling behavior
for the pion decay constant and pion masses. Furthermore the behavior of
these quantities with large $N_c$ does not agree with the expectation
from non-supersymmetric QCD. This outcome is not unexpected, since we
have no reliable information
about the K\"ahler potential and higher superderivative
terms in the effective action for supersymmetric QCD beyond weak coupling.
As our results illustrate, this lacking information is crucial for
any quantitative attempts to understand ordinary pion dynamics
in supersymmetric language. Although it may be possible to mimic the
correct decoupling behavior by appropriate rescalings of the terms in
the weak-coupling lagrangian, it is quite problematic to justify the
necessary terms. Thus the
apparent success associated with
finding the correct pattern of chiral symmetry breaking is of limited
direct applicability. However, one could use the exact supersymmetric
results even in the presence of supersymmetry breaking as a laboratory
to test calculational methods of strongly coupled theories~\cite{selipsky}.
Furthermore, lattice calculations can be compared with well-controlled
supersymmetric theories at strong coupling to gain insights into
both~\cite{ASPY,lattice}.

Supersymmetric QCD studies have resulted in a much improved understanding
of all supersymmetric field theories.  These advances have centered mainly
on the properties of the superpotential.  Together with the assumption
that the K\"ahler potential is non-singular in some region of field
space, this
allows us to understand the location of possible vacua and
massless
fields, but
does not yield complete information about massive states and
interactions. To learn more about strongly coupled
theories with large supersymmetry breaking will require further insights
and knowledge not only about the K\"ahler potential but also
about interactions with higher superderivatives, which contribute terms
to the potential with more than two powers of the auxiliary fields.  The fact
that we have almost no control over such contributions in $N=1$
theories in four dimensions seemingly
forbids progress in this direction.  Perhaps a more useful
starting point may be with a higher number of supersymmetries and/or
dimensions~\cite{mqcd} where the full lagrangian can be better controlled.

\noindent
{\it Acknowledgements.} 
We thank N.~Arkani-Hamed, M.~Dine and M.~Peskin for helpful discussions.
The work of SPM was supported in part by the US Department of Energy.


\end{document}